# Active Self-Tracking of Subjective Experience with a One-Button Wearable: A Case Study in Military PTSD

**Jakob Eg Larsen**
Technical University of Denmark
and TOTTI Labs
jaeg@dtu.dk

**Kasper Eskelund**
Danish Defence
Military Psychology Unit
VETC-MPA41@mil.dk

**Thomas Blomseth Christiansen**
Konsulent Blomseth
and TOTTI Labs
thomas@blomseth.dk

## ABSTRACT
We describe a case study with the participation of a Danish veteran suffering from post-traumatic stress disorder (PTSD). As part of psychotherapeutic treatment the participant and therapist have used our novel technique for instrumenting self-tracking of select aspects of subjective experience using a one-button wearable device. The instrumentation system is described along with the specific self-tracking protocol which defined the participant's self-tracking of a single symptom, namely the occurrences of a bodily experienced precursor to hyperarousal. Results from the case study demonstrate how self-tracking data on a single symptom collected by a patient can provide valuable input to the therapeutic process. Specifically, it facilitated identification of crucial details otherwise unavailable from the clinical assessment and even became decisive in disentangling different symptoms and their causes.

## INTRODUCTION
PTSD is a debilitating mental health disorder occurring after exposure to trauma such as interpersonal violence, motor vehicle accidents, sexual assault, or combat. It is characterized by pronounced heterogeneity of symptoms constellations [16, 6]. Since the 1990s, approx. 31,000 service members have deployed with the Danish armed forces to war zones in the Balkans, Iraq, and Afghanistan. Data from Danish ISAF (Afghanistan) veterans suggest that 9.7% of combat exposed soldiers will present with psychological symptoms at a level equivalent to a PTSD diagnosis 2.5 years after deployment [1].

Known evidence-based cognitive-behavioral psychotherapies for PTSD are effective for approximately 60% of treatment-seeking veterans [3]. A similar treatment response is found for pharmacological treatments [17]. These response rates highlight the need for innovative treatment proposals as well as a better understanding of lives lived with trauma.

In this paper we present a case study with a veteran who acquired PTSD symptoms after combat exposure in Afghanistan. In addition to receiving EEG-based neurofeedback as a treatment for hyperarousal [7, 12] at the Department of Military Psychology in the Danish Defence, the participant was self-tracking the occurrences of a precursor to one symptom (hyperarousal) in between sessions. Here we focus entirely on the self-tracking aspect. We describe the instrumentation used for self-tracking and data acquisition and discuss our findings from this case. In particular we discuss how frequent observations of a single symptom may jointly support treatment-seeking individuals and mental health professionals in identifying patterns that otherwise do not emerge as part of the usual psychotherapeutic treatment.

## INSTRUMENTING SUBJECTIVE EXPERIENCE SELF-TRACKING
Memory recall bias introduces issues with the reliability of self-reported data used as part of psychotherapeutic interventions. Whether such data is obtained from questionnaires, paper diaries, or as part of conversational therapy the problem remains that recall bias influences the accuracy of answers to questions, the ability to make accurate assessments, or even the ability to recall particular events at all.

Experience sampling methods (ESM) [15], ecological momentary assessment (EMA) [18], and ambulatory assessment (AA) [4] try to address the problem by collecting data closer to the situation and the moment where the phenomena of interest occur. Recently mobile phone text message systems, smartphone apps and most recently smartwatch apps [11, 8, 9] have become tools to acquire such data from individuals. However, these systems are typically driven by prompting the individual to report or momentarily assess an aspect as opposed to being driven by events of the individual's internal perception of their subjective experience.

While previous research has shown a greater accuracy of electronic diaries compared to paper diaries [10], we believe that electronic diaries must be taken a step further than current solutions in terms of simplicity, accuracy, and higher sampling rate. We present a self-developed data acquisition instrument that enables users to self-track select aspects of subjective experience by means of a wearable one-button device. The smartbutton can be mounted in a wristband so that it is easily accessible throughout daily life situations. As the device has only one button it can be operated without looking at it and due to the simplicity of the device an observation of a subjectively perceived phenomenon typically takes less than a second to make. Accordingly, it is fast and convenient for a user to register the occurrences of e.g. bodily sensations, feelings, emotions, thoughts, behaviors, and mental patterns in the moment, in the situation, and without involving memory. This is opposed to attempting to recall occurrences from memory when answering questionnaires, filling out a diary, using a smartphone app, or as part of a conversation post hoc.

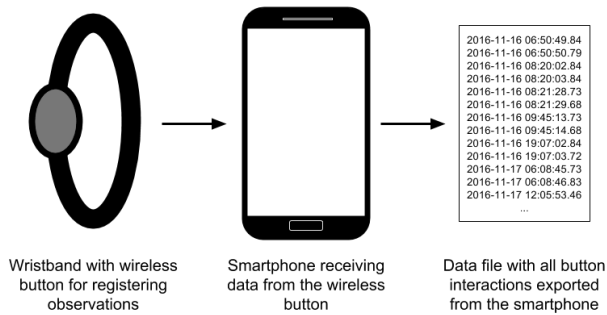

**Figure 1. The self-tracking instrument with a one-button wristband and smartphone for registering observations**

The instrumentation for active self-tracking of subjective experience (see Figure 1) contains a wristband with a Bluetooth enabled button. The wristband is carried by the self-tracking user, preferably at all times, and at the press of the button a signal is sent via a Bluetooth connection to a companion app running on a smartphone. In case there is no connection the button presses will be stored on the button and sent once a connection can be established. The smartphone app stores the data in the internal storage and allows the user to export a data file containing timestamps of all interactions made with the button. The system supports further annotation of observations including location information, if needed.

In the present case we focus on the one-button wearable instrument, whereas in previous work we have developed and carried out experiments with a smartwatch-based instrument also for *in situ* and *in momento* acquisition of data on subjective experience [13].

**CASE PN**
The therapist selected a candidate among participants at intake. The specific candidate participant was chosen on the criteria of a) a minimum of cognitive stability and b) an expectation of a sufficiently clearly defined PTSD case i.e. not involving severe social problems (such as homelessness) or severe substance abuse.

The participant (PN) is a 35-year-old male combat veteran with two deployments with the ISAF mission to Afghanistan and a history of depression, anxiety, and hyperarousal upon returning to civilian life. Before participating in the case study, PN had already completed two psychotherapeutic interventions totaling over 30 sessions. These cognitive-behavioral interventions succeeded in relieving depression and per psychoeducative instruction PN had learned efficient strategies for reducing anxiety. However, he was still suffering from seemingly unpredictable and uncontrollable peaks of arousal e.g. at times he needed to withdraw from his desk at work to achieve relaxation and prevent aggression and often refrained from social interaction with friends and relatives. Prolonged periods of hyperarousal impeded his ability to engage socially and hampered his efforts to achieve his stated goal of "living a normal, moderate life". As an attempt to reduce his reactiveness and arousal, he was referred to EEG neurofeedback treatment two months prior to the commencement of the study.

**METHOD**
Self-tracking was introduced at the first assessment session and the participant agreed to use the instrumentation from the second meeting. The participant was told he could stop self-tracking at any time without further consequences for his treatment or support from the Danish Veterans Centre.

The phenomenon to be tracked was defined at a two-hour assessment interview during which specific symptoms and their precursors were investigated with manualized cognitive-behavioral tools such as Cross Sectional Formulation, Five Column Thought Record [2]. The participant was able to identify a somatic marker that is a precursor of his hyperarousal. It was chosen from the outset and worked very well not necessitating alteration during the course of the study.

The participant and therapist defined an observation protocol and procedure for using the instrumentation: Two consecutive presses of the button when the participant experienced specific muscle contractions in the chest region and arms clearly preceding hyperarousal (in the case of PN, symptomatic readiness for fight and alertness towards non-present threats). The preceding bodily cue was chosen as to mark the stimuli and states preceding and perhaps causally linked to the symptom. Moreover, arrived in the state of hyperarousal, the participant had limited cognitive resources for registering the acute state. Two consecutive presses were chosen to be able to filter out false positives, i.e. mistakenly pressing the button once.

It was agreed that the participant would be carrying the smartbutton in a band on his left wrist under his sleeve making the button invisible to others, but still allowing presses to be delivered through the fabric of the shirt. In session or on request from the therapist the participant would export data from the smartphone app accompanying the smartbutton and send a file containing the collected button presses via email to the therapist for analysis.

**RESULTS**
After the initial assessment session the participant immediately started treatment with two to three therapeutic sessions per week and from week 7 into week 11 the participant had phone consultations with the therapist (see Figure 2).

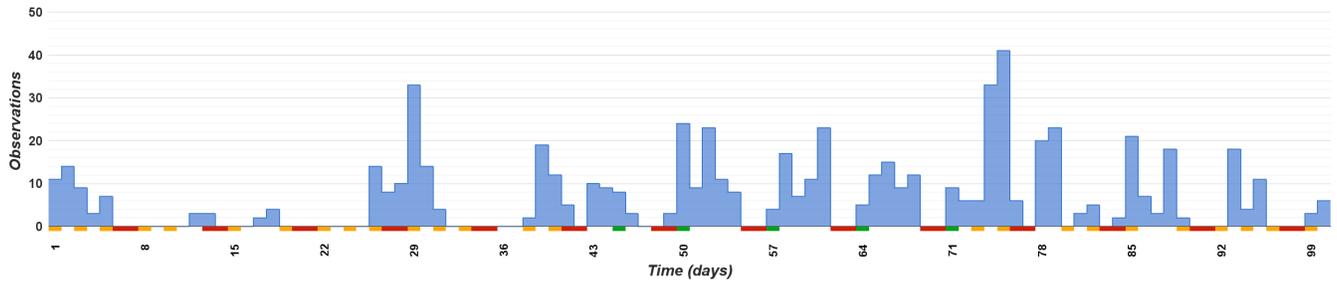

**Figure 2.** Observations (blue) as a function of time in the first 100 days of self-tracking as part of the treatment process. Events are indicated as follows: weekends (red), therapeutic sessions (yellow), and phone consultations (green)

| Hour of the day | 0 | 1 | 2 | 3 | 4 | 5 | 6 | 7 | 8 | 9 | 10 | 11 | 12 | 13 | 14 | 15 | 16 | 17 | 18 | 19 | 20 | 21 | 22 | 23 |
|---|---|---|---|---|---|---|---|---|---|---|---|---|---|---|---|---|---|---|---|---|---|---|---|---|
| Observations | 11 | 3 | 0 | 7 | 2 | 8 | 16 | 14 | 32 | 30 | 20 | 40 | 67 | 33 | 53 | 30 | 61 | 44 | 24 | 54 | 36 | 15 | 38 | 9 |
| Percentage | 1.7 | 0.5 | 0.0 | 1.1 | 0.3 | 1.2 | 2.5 | 2.2 | 4.9 | 4.6 | 3.1 | 6.2 | 10.4 | 5.1 | 8.2 | 4.6 | 9.4 | 6.8 | 3.7 | 8.3 | 5.6 | 2.3 | 5.9 | 1.4 |

**Table 1.** Distribution of observations on hours of the day and percentage of total observations. Noticeable is the uneven distribution of observations during daytime with afternoon hours having a higher frequency

The participant implemented the self-tracking protocol without hesitation and was actively pressing the button when observing the defined muscle contractions. During the first 100 days of self-tracking covered in this case study the participant made 647 observations of the phenomenon and went to 25 therapy sessions (see Figure 2). Due to a technical issue no data was collected during the second week i.e. the lack of observations was not due to lack of engagement by the participant.

The therapist plotted the collected data as a time series with occurrences of the muscle contraction as a function of time in order to look for patterns. Noticeable was a difference between weekdays and weekends i.e. a higher frequency of observations during weekdays (see Figure 2 and 3).

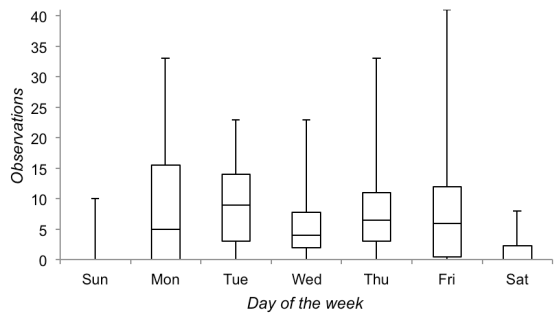

**Figure 3.** Box-and-Whisker plot of the distribution of observations on days of the week, indicating overall high variance and low frequency of observations during weekends

Furthermore it was identified that the frequency of observations generally was higher in the afternoon (see Table 1). In an investigational dialogue, participant and therapist jointly uncovered that the higher frequency in the afternoon on weekdays stemmed from stress induced by social interactions at work.

While regular assessment techniques had not been able to identify the occurrence of the precursory phenomenon at these particular times the self-tracking instrumentation enabled the participant to make observations in the situations they occurred. Thus self-tracking data on arousal offered a novel perspective for the participant on his own memory. Further, it enabled participant and therapist to jointly establish a probable causal link between specific stimuli and symptoms.

Specifically, self-tracking was effective in disentangling the main symptoms of hyperarousal, depression, and anxiety. Tracking was decisive in identifying that the three symptoms occurred in very different situations and were provoked by diverse circumstances. Hyperarousal occurred in social situations whereas depression and anxiety occurred at night when alone and feeling isolated. In turn, this allowed participant and therapist to pinpoint arousing vs. relieving behavior during everyday activities enabling a new dialogue on how to change stressful settings, or, how to cope with them if they could not be altered or avoided.

Seven weeks after the self-tracking had started the participant was denied disability compensation and went into acute crisis. Consequently, the therapeutic sessions were temporarily paused. Surprisingly, the participant kept wearing the instrumentation and continued self-tracking consistently even in the deteriorated situation (see Figure 2, days 43-72). During this period the frequency of symptoms went up but mainly within depression and anxiety. Only a slight increase in the frequency of observations was observed as the sensation tracked was not strongly related to these symptom dimensions (see Figure 2).

Conversely, the self-tracking data and the discussion it spawned has led the participant to reconsider continued education that could enable a career shift towards employment in a less stressful workplace. From a user engagement and systems design perspective we find it noticeable that at the time of writing the participant has been self-tracking using the instrumentation continuously for more than four months.

## DISCUSSION

The present case represents an initial exploration of the potential of the instrumentation and the data generated on subjective experience as the therapist only did a minimal integration into the treatment itself. Active self-tracking was used in a basic sense: Gathering data and discussing it to make an assessment of when and where arousal went up. In retrospect the therapist has noted that it could have been possible to engage in a deeper investigative dialogue based on the data.

A challenge in current clinical assessment techniques is the reliance on patient memory and the joint ability of therapist and patient to establish access to relevant information in gauging strength and characteristics of symptoms. In this state of affairs, however, the availability of information on symptoms in everyday life is likely to be impeded. Specifically, due to the effect of trauma on cognition and memory in PTSD, recallability of trauma-related thoughts and behaviors may be reduced. Trauma and related symptoms may further be associated with feelings of shame and guilt making them difficult to subject to clinical dialogue.

Psychotherapists are typically interested in the degree of suffering generated by a specific symptom. Thus they may ask patients questions like: "How often does this sensation occur?" Or: "How much does this feeling hinder you in doing what you want?" However, if the patient cannot differentiate between symptoms from memory, neither can the therapist. We believe that the present case study has demonstrated our proposition [14] that observational data on a single symptom collected with relatively high frequency by a patient can provide valuable input to the therapeutic process. We suggest that there is leverage to be found in the combination of what is meaningful to the patient, clinically relevant, and accurately self-reported.

The focus of the presented case has been entirely on active self-tracking with a wearable device. On the other hand, passive sensing of physiological markers such as heart rate and relevant behaviors such as physical activity, mobility [5] and social interactions [19] have received a lot of attention recently. Whereas PTSD-related symptoms such as social isolation and sleep are possible to infer from passive sensing data the mental states preceding and anticipating symptoms—bodily sensations, emotions, and thoughts—are not directly inferable from passive sensing data with current methods and instruments. Moreover, a change in a passive measure—e.g. the heart rate increasing during a state of anxiety—usually lags the somatic state when the individual has already passed the tipping point e.g. at a pulse of >120 bpm the threshold above which one would be able to change state by means of a cognitive strategy has already been passed. Only if the individual acknowledges a precursory phenomenon before the cascading autonomous nervous system processes kick in—i.e. before discernible signals from passively measureable phenomena are picked up—one will be able to not just survive one's anxiety but also to employ strategies to alter and endure it. We suggest that our instrumentation for active self-tracking complements passive sensing and may establish a novel way of producing ground truth for passive data towards recognizing precursory phenomena. Further, the inclusion of even a single, actively self-tracked phenomenon may yield the patient a sense of agency in the data gathering process that tendentially is absent in passive sensing alone.

Our instrumentation injects a new kind of data into the therapeutic process, however, we find it notable that the change which self-tracking effects in psychotherapy in terms of structure and content is minimal. Most of the healing mechanisms and interventions will—at least initially—remain unchanged. Only the accessibility to and quality of information about subjective experience change considerably. Additionally, the basic relation between patient and therapist will remain the same, although there is potential for becoming co-interpreters of the observational data, too, as opposed to being confined to being co-interpreters of the patient's utterances.

Finally, we consider if the scope of our proposed self-tracking with a one-button wearable instrument can be extended beyond a one-way observational tool to an interventional means of inducing and conditioning symptom-relieving behaviors like e.g. arousal-controlling breath techniques or techniques to interrupt rumination. This would add to the value of self-tracking by using the event of making an observation as an opportunity to invoke other basic symptom-relieving cognitive strategies.

## CONCLUSIONS

We have presented a novel instrumentation facilitating self-tracking of subjective experience with a one-button wearable device. Our instrument has been applied in a case study with a veteran suffering from PTSD. Between therapy sessions the participant self-tracked the occurrences of a single symptom (a precursor to hyperarousal) using the one-button wearable device. These *in situ* and *in momento* observations of a single symptom made by the participant were used by a therapist in the therapeutic process. In the case this type of symptoms data turned out to be valuable in understanding the triggers of hyperarousal. The additional level of detail became decisive in identifying crucial details that had not emerged from the conversational assessment. This case study suggests that frequently self-tracked data on everyday occurrences of symptoms can assist the therapist in obtaining an understanding of the mental disorder and symptoms otherwise inaccessible from regular clinical assessment.

## ACKNOWLEDGMENTS

We thank the participant in the case study for testing the self-tracking instrumentation and method.

**REFERENCES**
1. Andersen, S. B., Karstoft, K. I., Bertelsen, M., & Madsen, T. (2014). Latent trajectories of trauma symptoms and resilience: the 3-year longitudinal prospective USPER study of Danish veterans deployed in Afghanistan. The Journal of clinical psychiatry, 75(9), 1001-1008.
2. Beck, Aaron T., ed. (1979) Cognitive therapy of depression. Guilford press.
3. Bradley, R. (2005). A Multidimensional Meta-Analysis of Psychotherapy for PTSD. Am. J. Psychiatry *162*, 214
4. Conner, T. S., & Barrett, L. F. (2012). Trends in ambulatory self-report: the role of momentary experience in psychosomatic medicine. Psychosomatic medicine, 74(4), 327.
5. Cuttone, A., Lehmann, S., Larsen, J.E. (2014) Inferring Human Mobility from Sparse Low Accuracy Mobile Sensing Data. UbiComp 2014 3rd ACM Workshop on Computational Social Science.
6. Galatzer-Levy, I. R., & Bryant, R. A. (2013). 636,120 ways to have posttraumatic stress disorder. Perspectives on Psychological Science, 8(6), 651-662.
7. Eskelund, K., Karstoft, K.-I., Frost, M., Andersen, S.B. (2015). Latent Classes of PTSD Symptoms, Neural Activity and Neurocognitive Responses in Danish Treatment-Seeking Veterans. International Society for Traumatic Stress Studies. 31st Annual meeting, New Orleans 2015
8. Exler, A., Schankin, A., Klebsattel, C., & Beigl, M. (2016). A wearable system for mood assessment considering smartphone features and data from mobile ECGs. In Proceedings of the 2016 ACM International Joint Conference on Pervasive and Ubiquitous Computing: Adjunct (pp. 1153-1161). ACM.
9. Hänsel, K., Alomainy, A., & Haddadi, H. (2016). Large scale mood and stress self-assessments on a smartwatch. In Proceedings of the 2016 ACM International Joint Conference on Pervasive and Ubiquitous Computing: Adjunct (pp. 1180-1184). ACM.
10. Hufford, Michael R., Arthur A. Stone, Saul Shiffman, Joseph E. Schwartz, and Joan E. Broderick. "Paper vs. electronic diaries." Applied Clinical Trials 11, no. 8 (2002): 38-43.
11. Intille, S., Haynes, C., Maniar, D., Ponnada, A., & Manjourides, J. (2016). μEMA: Microinteraction-based ecological momentary assessment (EMA) using a smartwatch. In Proceedings of the 2016 ACM International Joint Conference on Pervasive and Ubiquitous Computing (pp. 1124-1128). ACM.
12. Karstoft, K.-I., Eskelund, K., Andersen, S.B. (2015). Latent Classes of PTSD Symptoms, Neurocognitive Responses and Heart Rate Variability in Danish Treatment-Seeking Veterans. International Society for Traumatic Stress Studies. 31st Annual meeting, New Orleans 2015
13. Larsen, J.E., Christiansen, T.B., Kaminski, T.R. (2016) Wearable, Digital Instrumentation for Better Quality Data on Subjective Experience in Health Research, Second Annual Health Data Exploration Network Meeting, San Diego.
14. Larsen, J.E., Eskelund, K., Christiansen, T.B. (2016) Keeping the Human in the Loop with Subjective Experience in Mental Health Tracking, Computing and Mental Health Workshop at CHI2016
15. Larson, R., & Csikszentmihalyi, M. (1983). The experience sampling method. New Directions for Methodology of Social and Behavioral Science, 15, 41-56.
16. Steenkamp, M. M., Boasso, A. M., Nash, W. P., Larson, J. L., Lubin, R. E., & Litz, B. T. (2015). PTSD symptom presentation across the deployment cycle. Journal of affective disorders, 176, 87-94.
17. Stein DJ, Ipser J, McAnda N. Pharmacotherapy of posttraumatic stress disorder: a review of meta-analyses and treatment guidelines. CNS Spectr. 2009 Jan;14(1 Suppl 1) 25-31. PMID: 19169191.
18. Stone, A. A., & Shiffman, S. (1994). Ecological momentary assessment (EMA) in behavorial medicine. Annals of Behavioral Medicine.
19. Stopczynski, A., Sekara, V., Sapiezynski, P., Cuttone, A., Madsen, M. M., Larsen, J. E., & Lehmann, S. (2014). Measuring large-scale social networks with high resolution. PloS one, 9(4), e95978